\documentclass[showpacs,amsmath,amssymb,prl,superscriptaddress,twocolumn]{revtex4}

\newcommand{\Lm}{\left\{}
\newcommand{\Rm}{\right\}}

\begin{document}

\title{$E_6$ grand unified theory with three generations from heterotic string}

\author{\quad Motoharu Ito}
\email{mot@eken.phys.nagoya-u.ac.jp}
\affiliation{Department of Physics, Nagoya University, Nagoya 464-8602, Japan}
\author{Shogo Kuwakino}
\email{skuwa@eken.phys.nagoya-u.ac.jp}
\affiliation{Department of Physics, Nagoya University, Nagoya 464-8602, Japan}
\author{Nobuhiro Maekawa}
\email{maekawa@eken.phys.nagoya-u.ac.jp}
\affiliation{Department of Physics, Nagoya University, Nagoya 464-8602, Japan}
\author{Sanefumi Moriyama}
\email{moriyama@math.nagoya-u.ac.jp}
\affiliation{Kobayashi Maskawa Institute, Nagoya University, Nagoya 464-8602, Japan}
\affiliation{Graduate School of Mathematics, Nagoya University, Nagoya 464-8602, Japan}
\author{Keijiro Takahashi}
\email{takahashi-2nyh@jp.nomura.com}
\affiliation{Department of Electrophysics, National Chiao-Tung University, Hsinchu, Taiwan, Republic of China}
\author{Kazuaki Takei}
\email{takei@eken.phys.nagoya-u.ac.jp}
\affiliation{Department of Physics, Nagoya University, Nagoya 464-8602, Japan}
\author{Shunsuke Teraguchi}
\email{teraguch@ifrec.osaka-u.ac.jp}
\affiliation{Department of Physics, Nagoya University, Nagoya 464-8602, Japan}
\affiliation{WPI Immunology Frontier Research Center, Osaka University, Osaka 565-0871, Japan}
\author{Toshifumi Yamashita}
\email{yamasita@eken.phys.nagoya-u.ac.jp}
\affiliation{Department of Physics, Nagoya University, Nagoya 464-8602, Japan}

\date{December 9, 2010}

\begin{abstract}
We construct two more supersymmetric $E_6$ grand unified models with
three generations within the framework of ${\mathbb Z}_{12}$ asymmetric
orbifold compactification of the heterotic string theory. 
Such an asymmetric orbifold is missing in the classification in the literature, 
 which concludes that only one $E_6$ model is possible. 
In both of the new models, an adjoint Higgs field is obtained in virtue of
the diagonal embedding method.
This method mods out the three $E_6$ factors of an even self-dual 
momentum-lattice by a permutation symmetry.
In order to realize the $(E_6)^3$ even self-dual lattice, we utilize 
the lattice engineering technique. 
Among the eight possible orbifold actions in our setup, two lead to new 
 $E_6$ models. 
Though these models still share the unsatisfactory issues with the known one,
our discovery raises hopes that excellent models that solve all
the problems in the supersymmetric grand unified models will be found 
in this framework.
\end{abstract}


\maketitle

\paragraph{Introduction.}
Superstring theory is one of the most promising candidates that 
describe quantum gravity and unify all the four fundamental forces of nature. 
Usually, however, superstring theory is defined in a ten-dimensional (10D) spacetime and 
its characteristic scale is taken to be the Planck scale, 
incredibly larger than that of 
the standard model (SM). 
If superstring theory truly describes 
our world, it must be an indispensable subject to find the way to the SM 
from superstring.
Unfortunately, superstring has a lot of perturbative vacua, and so far, 
 the way has not been established.
Therefore, it is worthwhile to ask phenomenological studies for hints.

Independently of the developments on superstring, 
 the supersymmetric (SUSY) grand unified theory (GUT)~\cite{GUT} is known 
 as an interesting candidate for the model 
beyond the SM. 
It unifies the three gauge groups $SU(3)_C\times SU(2)_L\times U(1)_Y$ in the
SM into a single gauge group. This unification is quantitatively supported by 
experiments which have revealed that the three gauge couplings in the 
 minimal supersymmetric SM meet 
 with a very good accuracy
 at a very high scale (the GUT scale)  close to the Planck scale. 
Moreover, it unifies one generation of quarks and leptons, which is dispersed 
among five multiplets in the SM, into one or two multiplets. This 
matter unification
is qualitatively supported by the measurements of quark/lepton masses and 
mixings:  
The pattern of the various hierarchical structures of 
 the masses and the mixings 
 can be explained by a simple assumption that 
 the hierarchies of the Yukawa couplings are induced mainly 
 by the ${\bf 10}$ multiplets of $SU(5)$~\cite{AB,Etwist,horizontal}. 

Among the SUSY-GUTs, the $E_6$ GUT~\cite{E6}, 
 which unifies all one generation
 quarks and leptons into a single ${\bf 27}$ multiplet,
has an advantage that the above
assumption for the Yukawa hierarchies, which must be 
made by hand in the $SU(5)$
unification, is naturally derived~\cite{Etwist,horizontal}. 
This advantage is particularly 
 important since it seems difficult to explain the hierarchical structure 
 in the minimal supersymmetric SM-like models obtained directly from superstring
\cite{heteroticMSSM,D-brane}.
Thus, apart from remaining issues, 
 such as the so-called doublet-triplet (DT) splitting problem and 
 the SUSY-flavor/CP problem, 
 it is plausible that this $E_6$ structure is realized. 

In addition, consistently with the $E_6$ structure, 
 it has been shown that the anomalous $U(1)_A$ gauge 
 symmetry~\cite{anomalousU(1)} 
 and the $SU(2)_H$ (or $SU(3)_H$) family symmetry~\cite{horizontal} 
 with a spontaneous CP violation~\cite{SCPV} respectively serve solutions to
the DT splitting problem and the SUSY-flavor/CP problem. 
Interestingly, both of the above two additional symmetries can be 
simultaneously adopted. 
Then, the resulting models are really promising, where 
almost all the phenomenological problems are solved, 
the realistic quark/lepton mass matrices are obtained naturally 
and all the three generations are unified into 
two multiplets (or a single one for $SU(3)_H$).

Thus, when we seek for the way to the SM from superstring, the above 
 scenario must be valuable to be considered, 
 though we should also keep 
 in mind the possibilities that not all of the 
 above additional symmetries, such as the anomalous $U(1)_A$ symmetry and 
 the $SU(2)_H$ symmetry, are actually realized. 
Therefore, 
 we assume only the four-dimensional (4D) SUSY $E_6$ unification in this letter. 
Namely,
 our strategy is 
 to construct phenomenological string models with the following 
 minimal requirements, 
\begin{itemize}
\item $E_6$ unification group,
\item 4D ${\cal N}=1$ SUSY,
\item adjoint Higgs field,
\item three families,
\end{itemize}
 anticipating to find models with all or some of the additional symmetries 
 mentioned above. 

In the literature, despite decades of research, 
 only one model with these requirements has been reported~\cite{KTE6}. 
The authors of the reference claimed that they classified the models with the minimal 
 requirements and a hidden non-Abelian gauge symmetry as well which may be 
 useful to break the SUSY dynamically. 
Unfortunately, however, the above-mentioned 
 additional symmetries are not realized in their model. 
Therefore, we would like to search for more 4D ${\cal N}=1$ SUSY $E_6$
models in string theory, by relaxing the requirement of the hidden non-Abelian gauge 
 symmetry as it is possible to break the SUSY in other ways 
 (for example, in a meta-stable vacuum~\cite{ISS, SUSYbr}).

\paragraph{Strategy.}

Let us work on the heterotic string theory~\cite{hetero}, 
 where the $E_6$ unified group can be realized  without much difficulty. 
In contrast to the F-theory~\cite{F-theory} which is another related 
 framework realizing the $E_6$ group~\cite{F-E6} and has attracted attention 
 recently~\cite{F-GUT}, the heterotic string theory has a microscopic 
 description by a Lagrangian and thus any quantity is, at least in principle, 
 calculable. 

In the heterotic string, to obtain the 4D effective theory, 
 twenty-two extra dimensions in the 
 left-moving bosonic string and six in the right-moving superstring 
 should be compactified. 
Then, both the left-moving and the right-moving momenta are quantized and 
 compose a lattice. 
{}For the consistency of string theory, 
 the partition function, 
 namely, the one-loop vacuum diagram of the closed string
must be invariant under the modular transformations
 ${\mathcal T}:\tau\to\tau+1$ and ${\mathcal S}:\tau\to-1/\tau$, 
 where $\tau$ is the moduli of the worldsheet torus.
This requires that the lattice is even and self-dual 
 with a $(22,6)$-Lorentzian signature.
A spacetime gauge symmetry $G$ (in particular $E_6$) is realized 
 when this momentum-lattice contains the 
 appropriately normalized Lie lattice of $G$ in the left-moving part.

To reduce the 4D ${\cal N}=4$ SUSY to ${\cal N}=1$, six components among eight
 of the massless 10D spinor mode 
 in the right-moving superstring have to be projected out. 
This is achieved  
 by an orbifold compactification~\cite{orbifold}, 
 which identifies the compactified space under an action of a point group that 
 leaves the lattice unchanged. 
In particular, the compactified right-moving six dimensions should be fully rotated 
 with three nontrivial rotating angles $t_{Ri}$ that satisfy $\sum_{i=1}^3t_{Ri}=0$ 
 (mod $2\times2\pi$).

In general, when the heterotic string realizes a spacetime gauge
symmetry, the currents of the corresponding worldsheet theory form
a Kac-Moody algebra: 
$[j^a_m, j^b_n]=if^{ab}{}_cj^c_{m+n}+km\delta^{ab}\delta_{m+n,0}$.
Here, $j^a_m$ is the Laurent coefficient of the worldsheet current
$j^a(z)=\sum_{m\in{\mathbb Z}}j^a_mz^{-m-1}$, $f^{ab}{}_c$ is a structure constant 
and the integer $k$ is a Kac-Moody level.
The usual orbifold construction described above 
 realizes the lowest Kac-Moody level ($k=1$), 
 while it is known~\cite{adjoint} 
 that a higher level is necessary 
 to obtain adjoint Higgs fields. 
A way to increase the level is the so-called diagonal embedding 
 method~\cite{DiagonalEmbedding}, 
 where $K$-copies of the current $(j)_I$ with 
 level $k=1$ are permuted by an orbifold action so that only the diagonal part 
 $j_{\rm diag}=\sum_{I=1}^K(j)_I$ remains phaseless under the action. 
It is easy to see that $j_{\rm diag}$ satisfies the Kac-Moody algebra with $k=K$. 
The other eigenstates have nontrivial phases, and thus do not contribute to
 the 4D gauge 
 multiplets, while some of them may couple with chiral multiplets 
 in the right-mover to cancel the phases, resulting in adjoint Higgs fields.
It is also possible that adjoint Higgs fields appear in twisted sectors.

Unfortunately, it is not easy to clarify the condition in string theory 
 to construct models with three generations, 
 while there is a conjecture that the number of the generations is proportional 
 to the Kac-Moody level~\cite{KTE6}. 
Therefore, we start with the construction of 4D ${\cal N}=1$ SUSY $E_6$ 
 models with an adjoint Higgs field, leaving the numbers of generations to be 
 determined model-by-model.
To summarize, we take the following strategy:
\begin{enumerate}
\item we prepare a (22,6)-dimensional even self-dual lattice 
      with equivalent $K$-copies of the left-moving $E_6$ lattice, 
\item we consider an orbifold identification that includes
 \begin{enumerate}
 \item a permutation among the $E_6$ factors,
 \item rotations of the right-moving  six dimensions with three nonzero 
       angles $t_{Ri}$ satisfying $\sum_{i=1}^3 t_{Ri}=0$ (mod $2\times2\pi$),
 \end{enumerate}
\item we find out the number of the generations. 
\end{enumerate}

According to the conjecture, $k=3$ is needed for the three generations, 
 and we take this choice hereafter. 
In this case, the left-moving $(E_6)^3$ lattice occupies 18 dimensions 
 and cannot be fitted in the  16 extra dimensions (with respect to the 10D 
 viewpoint). 
This means that the usual left-right symmetric treatment of the six 
 extra dimensions is not valid and we have to work in the asymmetric 
 orbifold~\cite{asymmetric} 
 with a Narain compactification~\cite{Narain}. 
In contrast to the symmetric orbifold, the general rules for consistent 
 models are rather involved in the asymmetric orbifold~\cite{KTE6}, 
 and thus, we calculate the one-loop partition functions explicitly to check 
 the modular invariance (see Ref.~\cite{paper} for the details).

\paragraph{Setup.}

The lattice engineering technique \cite{LSW} is helpful in
constructing desired lattices.
The essence of the technique is that a lattice (for example, the $A_2$
lattice) transforms oppositely as its complement lattice in the
Euclidean even self-dual $E_8$ lattice (the $E_6$ lattice for the
$A_2$ example) under the modular transformation.
Thanks to this, we can always replace the left-moving $A_2$ lattice by the
right-moving $\bar E_6$ lattice (denoted with a bar) and vice versa.
Subsequently, since the $\bar E_6$ lattice is decomposed into three
$\bar A_2$ lattices, we
can construct a left-moving $(E_6)^3$ lattice out of the right-moving
$(\bar A_2)^3$ lattice using the same technique again.
Thus, we can easily convert an $A_2$ lattice into three equivalent 
 $E_6$ lattices.

With all these insights in mind, let us pick up the
$E_6\times\bar E_6$ lattice as our starting point.
After decomposing $E_6$ into $(A_2)^3$, we end up with
$[(A_2)^2\times(E_6)^3]\times\bar E_6$ using the above technique.
Though this lattice is the same as the one used in Ref.~\cite{KTE6} 
 constructed from the 10D $SO(32)$ heterotic string
 by a compactification with Wilson lines, 
 our method is more direct and hence many Narain 
 lattices with $(E_6)^3$ symmetry are accessible with their discrete
 symmetries manifest.

The orbifold action on the three left-moving $E_6$ factors 
 is chosen to be a permutation among them.
It turns out that a shift along the diagonal factor 
 has to be introduced in addition, as a source of the asymmetry 
 between the numbers of the generations and the antigenerations
 in the twisted sectors.

A natural candidate for the action on the right-moving factor is 
 the rotation by the Coxeter element of the right-moving $\bar E_6$ lattice, 
 which is an element of a point group ${\mathbb Z}_{12}={\mathbb Z}_3\times {\mathbb Z}_4$. 
Though it was claimed that ${\mathbb Z}_{2}$ is the only possible symmetry to add to 
 the ${\mathbb Z}_3$ symmetry for the above permutation~\cite{KTE6}, 
 we find no reasons to exclude this possibility. 
Thus, we choose this rotation, which corresponds to the one with three 
 angles $t_{R}=2\pi(1,\,4,\,-5)/12$ 
 classified as ${\mathbb Z}_{12}$-I~\cite{Z12}, 
 as part of our setup. 

Then, the remaining options are actions on the two left-moving $A_2$ factors. 
The allowed choices on each factor, labeled by $i$, are 
\begin{itemize}
\item shift $s_{Li}$, with $12s_{Li}=0$ (mod roots),
\item rotation with an angle $t_{Li}=2\pi/3$, 
\item Weyl reflection. 
\end{itemize}
There are a lot of choices of $s_{Li}$ but many of them are related  
 to each other by transformations under the symmetry of the 
 $A_2$ lattice, leading to identical models.
In addition,
 the modular invariance does not allow arbitrary choices, 
 but only certain combinations. 
Thus, there remain only a few possible actions: 
\begin{eqnarray*}
&
 \Lm {(2,0)},\, {(4,0)},\, ``{\rm rot}"\Rm \otimes 
    \Lm {(0,0)},\, {(6,0)} \Rm, 
&\\&
 (1,0)\otimes (3,6),\quad
 (1,6)\otimes (3,0),
&
\end{eqnarray*}
 where $``{\rm rot}"$ denotes the $1/3$ rotation while $(n,m)$ 
 represents the shift defined by 
 the vector $s=(n\alpha_1+m\alpha_2)/12$, with $\alpha_i$ being 
 the simple roots of $A_2$. 
In the first line, we have three options for the action on one of $A_2$ lattices, 
 and two for the other. 
Therefore, there are, in total,  $3\times2+1\times1+1\times1=8$ consistent models 
possible in this setup. 
Note that the order is irrelevant since the two $A_2$ factors are equivalent.

\paragraph{Models with three generations.}

Once fixing the orbifold action, one can calculate the partition function 
 (see, for example, Refs.~\cite{KTE6,paper}), 
 which contains information of the spectrum of 
 the model. 
It turns out that, among the above eight models, 
 three lead to vanishing net generation numbers, 
 while other two and the remaining three, respectively, have nine and three
 net generations. 
Here, we concentrate on the last three, 
\begin{equation*}
 (2,0) \otimes (6,0),\quad
 (1,0) \otimes (3,6), \quad
 (1,6) \otimes (3,0),
\end{equation*}
and we call them Model 1, 2, 3, respectively.

Model 1 and Model 2 have the gauge group $E_6 \times SU(2) \times U(1)^3$, 
 and Model 3 has $E_6 \times U(1)^4$.
Their massless spectra are listed in TABLE~\ref{spectra}.
We find five generations and two antigenerations in Model 1 
 and Model 3, while four and one in Model 2. 
Thus we obtain three models with three generations. 
Each model contains an $E_6$ adjoint Higgs field in the untwisted sector.

Model 1 results in the same massless spectrum as the ${\mathbb Z}_6$ model 
 in Ref.~\cite{KTE6}. 
The other two, Model 2 and Model 3, are new. 
Model~3 does not contain any hidden 
 non-Abelian gauge symmetry, which is one of the requirements 
 of the classification in Ref.~\cite{KTE6}, while Model 1 and Model~2 do.
We also find that these models have  
 $({\mathbb Z}_3)^3$ symmetry which remains unbroken even after all the singlets 
 develop nonvanishing vacuum expectation values and, 
 unfortunately, do not possess the additional 
 symmetries~\cite{horizontal,anomalousU(1),SCPV}.
Thus, the traditional SUSY-GUT problems, such as the DT splitting
problem and the SUSY-flavor/CP problem,
are not resolved in these models.

\begin{table*}
\caption[smallcaption]{The massless spectra of the models with three generations:  
$U$ and $T_\alpha$ denote the untwisted and various twisted sectors, respectively. 
The quantum numbers of left-handed chiral multiplets and the normalization 
of the $U(1)$ charges are shown.
The gravity and gauge multiplets are omitted.
}
\label{spectra}
\renewcommand{\arraystretch}{1.2}
\begin{tabular}{ccccc}
\hline\hline
          & Model 1 & Model 2 & Model 3  \\ 
\hline
\quad\phantom{,} gauge symmetry \quad\phantom{,}
    & \phantom{,} $E_6 \times SU(2) \times U(1)^3 $ \phantom{,} 
    & \phantom{,} $E_6 \times SU(2) \times U(1)^3 $ \phantom{,} 
    & $E_6 \times U(1)^4$ \\
\hline
$U$ & \begin{tabular}{c}$({\bf 1} ,{\bf 1}, +6,0,0)_{{\rm L}}$ \\
                        $( {\bf 78}, {\bf 1},0,0,0  )_{{\rm L}}$ 
      \end{tabular}
    & \begin{tabular}{c}$( {\bf 1},{\bf 1},+6,\pm 3, 0 )_{{\rm L}}$ \\
                        $( {\bf 78},{\bf 1},0,0,0 )_{{\rm L}}$ 
      \end{tabular} 
    & \begin{tabular}{c}$({\bf 1},-6,0,0,0)_{{\rm L}}$ \\
                        $({\bf 1},+3,\pm 6,0,0)_{{\rm L}}$ \\
                        $({\bf 78},0,0,0,0)_{{\rm L}}$
      \end{tabular}
\\ \phantom{,}\vspace{-4.3mm} \\
$T_1$ & $( {\bf 27},{\bf 1},+1,0,\pm1 )_{{\rm L}}$ 
      & --- 
      & \phantom{,} $({\bf 27},-1,-1,+1,0)_{{\rm L}}$ \phantom{,}  
\\ \phantom{,}\vspace{-4.3mm} \\
$T_2$ & $( \overline{{\bf 27}}, {\bf 1},-1,\pm1,0 )_{{\rm L}}$ 
      & $( \overline{{\bf 27}},{\bf 1},+2,0,-2 )_{{\rm L}}$ 
      &  $(\overline{{\bf 27}},+1,0,0,\pm1)_{{\rm L}}$   
\\ \phantom{,}\vspace{-4.3mm} \\
$T_3$ & \begin{tabular}{c} $ 2 ( {\bf 1},{\bf 1},-3,0,\pm3  )_{{\rm L}}$ 
        \end{tabular}
      & \begin{tabular}{c} $( {\bf 1},{\bf 1},-3,\pm 3,-3 )_{{\rm L}}$ 
        \end{tabular}
      & \begin{tabular}{c} $({\bf 1},+3,-3,+3,0)_{{\rm L}}$  \\ 
                           $({\bf 1},+3,+3,-3,0)_{{\rm L}}$
        \end{tabular}
\\ \phantom{,}\vspace{-4.3mm} \\
$T_4$ & \begin{tabular}{c} $( {\bf 27},{\bf 1},-2,0,0  )_{{\rm L}}$
        \end{tabular}
      & \begin{tabular}{c} $( {\bf 27},{\bf 1},-2,\pm 1,0 )_{{\rm L}}$
        \end{tabular}
      & \begin{tabular}{c} $({\bf 27},+2,0,0,0)_{{\rm L}}$ \\
                           $({\bf 27},-1,\pm2,0,0)_{{\rm L}}$
        \end{tabular}
\\ \phantom{,}\vspace{-4.3mm} \\
$T_5$ & $( {\bf 27},{\bf 1},+1,0,\pm1 )_{{\rm L}}$ 
      & $( {\bf 27},{\bf 1},+1,\pm 1,+1 )_{{\rm L}}$ 
      &  $({\bf 27},-1,+1,-1,0)_{{\rm L}}$  
\\ \phantom{,}\vspace{-4.3mm} \\
$T_6$ & \begin{tabular}{c} $( {\bf 1},{\bf 2},0,0,\pm3 )_{{\rm L}}$ \\
                           $( {\bf 1},{\bf 1},+3,\pm3,0 )_{{\rm L}}$
        \end{tabular}
      & \begin{tabular}{c} $( {\bf 1},{\bf 2},0,\pm 3,0 )_{{\rm L}}$ \\
                           $( {\bf 1},{\bf 1},-6,0,+6 )_{{\rm L}}$
        \end{tabular}
      & \begin{tabular}{c} $({\bf 1},-3,0,0,\pm3)_{{\rm L}}$ \\
                           $({\bf 1},0,+6,-2,0)_{{\rm L}}$ \\
                           $({\bf 1},0,-6,+2,0)_{{\rm L}}$
        \end{tabular} \\
\hline
normalization of U(1) & $(\frac{\sqrt{2}}{6}, \frac{\sqrt{6}}{6}, \frac{\sqrt{6}}{6} )$        & $(\frac{\sqrt{2}}{12}, \frac{\sqrt{6}}{6}, \frac{\sqrt{6}}{12} )$        & $(\frac{\sqrt{2}}{6}, \frac{\sqrt{6}}{12}, \frac{\sqrt{2}}{4}, \frac{\sqrt{6}}{6})$        \\
\hline\hline
\end{tabular}
\end{table*} 
%

\paragraph{Summary.}

In this letter, 
 we construct 4D SUSY level-$3$ $E_6$ models. 
The $k=3$ $E_6$ gauge symmetry is realized from three copies of $k=1$ $E_6$ 
 symmetry via the diagonal embedding. 
We utilize the lattice engineering technique, instead of the compactification of 
 the usual 10D heterotic string models with Wilson lines, 
 to construct Narain lattices containing three copies of the $E_6$ lattices. 
This technique allows us to construct 
 new even self-dual lattices from a known one 
 in a simple way, and thus, makes it easier to access new models.
Though here we work only on the same lattice as the one studied in Ref.~\cite{KTE6}  
 where the lattice is obtained through Wilson lines, 
 we show that Narain lattices with desired three copies of $E_6$ 
can be immediately constructed from any lattice containing $A_2$.

Then, we examine all the possible ${\mathbb Z}_{12}$ actions which 
 are missing in the classification in the literature~\cite{KTE6}, and we find 
 three models with the minimal requirements. 
One of them has the same spectrum as the model~\cite{KTE6} 
 that has been the only one proposed so far. 
The other two are new. 
While one does not have any hidden non-Abelian gauge symmetry, 
 the other does, and thus should be added into the classification. 

The two new models contain neither an 
 $SU(2)_H$ family symmetry nor an 
 anomalous $U(1)_A$ gauge symmetry
 which make the $E_6$ models more attractive. 
Given that we have shown
 there are $E_6$ models besides  
 the unique one proposed so far, it is 
 worthwhile to look for more $E_6$ models, especially 
 the excellent models with the above additional symmetries. 
{}For this purpose, our systematic construction of the $E_6$ models 
 will be useful.

\paragraph{Acknowledgments.}

We appreciate Y.~Kawamura, K.~Hosomichi, H.~Kanno, T.~Kobayashi, J.~C.~Lee,
S.~Mizoguchi, Y.~Sugawara for valuable discussions. 
S.M.\ would like to thank Yukawa Institute for hospitality. 
This work was partially supported by the Grant-in-Aid for Nagoya
University Global COE Program (G07), 
by MEXT of Japan [Grant No. 22011004 (N.M.),  Grant No. 21740176 (S.M.)] 
and by JSPS (T.Y.).


\begin{thebibliography}{99}
%
\bibitem{GUT}
  H.~Georgi and S.~L.~Glashow,
  Phys.\ Rev.\ Lett.\  {\bf 32} (1974) 438.
%
%
\bibitem{AB}
  C.~H.~Albright, K.~S.~Babu and S.~M.~Barr,
  Phys.\ Rev.\ Lett.\  {\bf 81} (1998) 1167;
%
  W.~Buchmuller and T.~Yanagida,
  Phys.\ Lett.\  B {\bf 445} (1999) 399;
%
  Q.~Shafi and Z.~Tavartkiladze,
  Phys.\ Lett.\  B {\bf 451} (1999) 129.
%
%
\bibitem{Etwist}
  J.~Sato and T.~Yanagida,
  Phys.\ Lett.\  B {\bf 430} (1998) 127;
%
  M.~Bando and T.~Kugo,
  Prog.\ Theor.\ Phys.\  {\bf 101}, 1313 (1999);
%
  M.~Bando and N.~Maekawa,
  Prog.\ Theor.\ Phys.\  {\bf 106} (2001) 1255.
%
%
\bibitem{horizontal}
  N.~Maekawa,
  Phys.\ Lett.\  B {\bf 561}, 273 (2003);
%
  Prog.\ Theor.\ Phys.\  {\bf 112}, 639 (2004).
%
%
\bibitem{E6}
  F.~Gursey, P.~Ramond and P.~Sikivie,
  Phys.\ Lett.\  B {\bf 60} (1976) 177.
%
%
\bibitem{heteroticMSSM}
  W.~Buchmuller, K.~Hamaguchi, O.~Lebedev and M.~Ratz,
  Phys.\ Rev.\ Lett.\  {\bf 96} (2006) 121602;
%
  Nucl.\ Phys.\  B {\bf 785} (2007) 149.
%
%
\bibitem{D-brane}
 G.~Honecker and T.~Ott,
  Phys.\ Rev.\  D {\bf 70} (2004) 126010
  [Erratum-ibid.\  D {\bf 71} (2005) 069902].
%
%
\bibitem{anomalousU(1)}
  N.~Maekawa,
  Prog.\ Theor.\ Phys.\  {\bf 106} (2001) 401;
%
  N.~Maekawa and T.~Yamashita,
  Prog.\ Theor.\ Phys.\  {\bf 107} (2002) 1201;
  Prog.\ Theor.\ Phys.\  {\bf 110} (2003) 93.
%
%
\bibitem{SCPV}
  M.~Ishiduki, S.~G.~Kim, N.~Maekawa and K.~Sakurai,
  Phys.\ Rev.\  D {\bf 80}, 115011 (2009)
  [Erratum-ibid.\  D {\bf 81}, 039901 (2010)];
  H.~Kawase and N.~Maekawa,
  Prog.\ Theor.\ Phys.\  {\bf 123}, 941 (2010).
%
%
\bibitem{KTE6}
  Z.~Kakushadze and S.~H.~H.~Tye,
  Phys.\ Rev.\  D {\bf 54} (1996) 7520;
%
  Phys.\ Rev.\  D {\bf 55}, 7878 (1997).
%
%
\bibitem{ISS}
  K.~A.~Intriligator, N.~Seiberg and D.~Shih,
  JHEP {\bf 0604} (2006) 021.
%
%
\bibitem{SUSYbr}
  S.~G.~Kim, N.~Maekawa, H.~Nishino and K.~Sakurai,
  Phys.\ Rev.\  D {\bf 79} (2009) 055009.
%
%
\bibitem{hetero}
  D.~J.~Gross, J.~A.~Harvey, E.~J.~Martinec and R.~Rohm,
  Phys.\ Rev.\ Lett.\  {\bf 54} (1985) 502;
%
  Nucl.\ Phys.\  B {\bf 256} (1985) 253;
%
  Nucl.\ Phys.\  B {\bf 267} (1986) 75.
%
%
\bibitem{F-theory}
  C.~Vafa,
  Nucl.\ Phys.\  B {\bf 469} (1996) 403.
%
%
\bibitem{F-E6}
  C.~M.~Chen and Y.~C.~Chung,
  arXiv:1010.5536 [hep-th].
%
%
\bibitem{F-GUT}
  R.~Donagi and M.~Wijnholt,
  arXiv:0802.2969 [hep-th];
%
  C.~Beasley, J.~J.~Heckman and C.~Vafa,
  JHEP {\bf 0901} (2009) 058.
%
%
\bibitem{orbifold}
  L.~J.~Dixon, J.~A.~Harvey, C.~Vafa and E.~Witten,
  Nucl.\ Phys.\  B {\bf 261} (1985) 678;
%
  Nucl.\ Phys.\  B {\bf 274} (1986) 285.
%
%
\bibitem{adjoint}
  D.~C.~Lewellen,
  Nucl.\ Phys.\  B {\bf 337} (1990) 61.
%
%
\bibitem{DiagonalEmbedding}
  K.~R.~Dienes and J.~March-Russell,
  Nucl.\ Phys.\  B {\bf 479} (1996) 113.
%
%
\bibitem{asymmetric}
  K.~S.~Narain, M.~H.~Sarmadi and C.~Vafa,
  Nucl.\ Phys.\  B {\bf 288} (1987) 551.
%
%
\bibitem{Narain}
  K.~S.~Narain,
  Phys.\ Lett.\  B {\bf 169}, 41 (1986).
%
%
\bibitem{paper}
  M.~Ito, S.~Kuwakino, N.~Maekawa, S.~Moriyama, K.~Takahashi,
  K.~Takei, S.~Teraguchi and T.~Yamashita,
  arXiv:1104.0765 [hep-th].
%
%
\bibitem{LSW}
  W.~Lerche, A.~N.~Schellekens and N.~P.~Warner,
  Phys.\ Rept.\  {\bf 177}, 1 (1989).
%
%
\bibitem{Z12}
 Y.~Katsuki, Y.~Kawamura, T.~Kobayashi, N.~Ohtsubo, Y.~Ono and K.~Tanioka,
  Nucl.\ Phys.\  B {\bf 341} (1990) 611.
%
%
\end{thebibliography}
\end{document}